\documentclass[pre,showpacs]{revtex4}
\usepackage{graphicx}
\usepackage{amsmath}
\usepackage{amssymb}
\usepackage{psfrag}
\usepackage{bm}
\usepackage{times}

\begin{document}
\date{\today}
\title{Lateral diffusion of a protein on a fluctuating membrane}
\author{E. Reister}
\email[Electronic address: ]{reister@theo2.physik.uni-stuttgart.de}
\author{U. Seifert}
\affiliation{
II. Institut f\"ur Theoretische Physik, Universit\"at Stuttgart - 70550 Stuttgart, Germany
}

\begin{abstract}
  Measurements of lateral diffusion of proteins in a membrane typically assume
  that the movement of the protein occurs in a flat plane.  Real membranes,
  however, are subject to thermal fluctuations, leading to movement of an
  inclusion into the third dimension.  We calculate the magnitude of this
  effect by projecting real three-dimensional diffusion onto an effective one
  on a flat plane. We consider both a protein that is free to diffuse in the
  membrane and one that also couples to the local curvature. For a freely
  diffusing inclusion the measured projected diffusion constant is up to 15\%
  smaller than the actual value. Coupling to the
  curvature enhances diffusion significantly up to a factor of two. 
\end{abstract}
\maketitle
\section*{Introduction}
During the last two decades significant progress has been made regarding novel
microscopy techniques, that are mainly based on the observation of single or
few molecules. With the aid of single particle tracking, photonic force
microscopy, or fluorescence-based single-molecule microscopy one is now
capable of studying the dynamic properties of lipids or proteins in membranes
with positional accuracies smaller than $40\ensuremath{\unskip\,\text{nm}}$ and a time resolution that
can be as small as tens of microseconds, see the review \cite{Lommerse} and
references therein.  These techniques together with other methods like
fluorescence correlation spectroscopy, reviewed in~\cite{Schwille}, have made
it possible to measure lateral diffusion of single molecules in a membrane
very accurately giving insight into the organisation of biomembranes. Although
these high resolution methods have become standard practice, it is typically
overlooked that a membrane is soft and therefore subject to thermal shape
undulations.  Diffusion coefficients extracted from experimental data usually
correspond to projected diffusion in a flat plane. The fluctuations of the
membrane, however, lead to a three-dimensional motion of the diffusing
protein, but unlike free three-dimensional diffusion the particle is confined
to the membrane.  Since the fluctuation spectrum of the membrane can be
affected by external parameters, like e.g.~temperature, osmotic pressure or pH
differences~\cite{Seifert,Gompper}, these high accuracy methods should make
the apparent change in diffusion experimentally measurable.

In this paper, we are interested in how big the difference between the actual
intramembrane and the measured projected diffusion constant is. We study both
the case of a protein, which is free to diffuse~\cite{Aizenbud,Gustaffson},
and a curvature-coupled protein with (or inducing) a spontaneous curvature.
Most studies of membrane-inclusions within various models, which include rigid
inclusions of various shapes, or proteins with a large domain outside the
membrane, analyse the static interaction with the membrane and/or the
interaction between inclusions~\cite{Weikl_2,Kim,Netz,Dan,Goulian,Lipowsky},
and have not yet addressed the dynamical issue of diffusion.
\section*{Free diffusion}
First, we regard the simplest case of a protein free to diffuse in the
membrane. We assume that the diffusion is not influenced by the local shape of
the membrane, therefore apart from the inclusion being confined to the
membrane there are no interactions between membrane and protein.  To describe
diffusion on a curved surface, the Laplace operator in the diffusion equation
for a plane needs to be replaced by the Laplace-Beltrami operator. If the
position of the surface $\tilde{\mathbf{r}}$ is expressed in the Monge
representation, i.e.~$\tilde{\mathbf{r}}=(x,y,h(x,y))$, the resulting
Smolouchovski equation is~\cite{Aizenbud}
\begin{equation}
  \begin{split}
    \frac{\partial P(x,y,t)}{\partial
      t}=D\frac{1}{g}&\left\{(1+h_y^2)\frac{\partial^2 P}{\partial
        x^2}+(1+h_x^2)\frac{\partial^2 P}{\partial y^2}-2h_x h_y \frac{\partial^2
        P}{\partial x\partial y}\right.\\
    -\frac{1}{g}&\left[h_{xx}h_x(1+h_y^2)+h_{yy}h_x(1+h_x^2)-2h_{xy}h_y
      h_x^2\right]\frac{\partial P}{\partial x}\\
    -\frac{1}{g}&\left.\left[h_{yy}h_y(1+h_x^2)+h_{xx}h_y(1+h_y^2)-2h_{xy}h_x
        h_y^2\right]\frac{\partial P}{\partial y}\right\},
  \end{split}
\end{equation}
with $h_x\equiv\partial h(x,y)/\partial x$ --for the other subscripts
accordingly-- and $g\equiv 1+h_x^2+h_y^2$. $D$ is the diffusion constant on
the curved surface, $P(x,y,t)dxdy$ is the probability to find the diffusing
particle in the area element $dxdy$ at point $\tilde{\mathbf{r}}$.  The
membrane is subject to thermal fluctuations. Similarly to the calculation of
diffusion within the Zimm model~\cite{Doi} we introduce a preaveraging
approximation: Instead of using the prefactors, that contain $h_x$, $h_y$,
$h_{xx}$, $h_{yy}$, $h_{xy}$, and therefore explicitly depend upon position
and time, we replace them by their thermal averages $\langle\ldots\rangle$.
This approximation is valid if the time scale of protein diffusion is much
larger than that of membrane shape fluctuations. Due to their asymmetry
averages like $\langle h_xh_y/g\rangle$ and many others vanish.
Only $\langle h_x^2/g\rangle=\langle h_y^2/g\rangle$ are non-zero
contributions.  This considerably simplifies the diffusion equation
\begin{equation}
  \frac{\partial P(x,y,t)}{\partial
    t}=D\left\{\left(1-\left\langle\frac{h_x^2}{g}\right\rangle\right)
\frac{\partial^2 P}{\partial x^2}+\left(1-\left\langle\frac{h_y^2}{g}
\right\rangle\right)\frac{\partial^2 P}{\partial y^2}\right\}.
\label{eq:diff_compact}
\end{equation}
Average quantities of an isotropic membrane cannot be different for the $x$-
or $y$-direction, and the effective diffusion constant $D_{\text{proj}}$, that
would be measured in the $x$-$y$-plane is rescaled to
\begin{equation}
  \frac{D_{\text{proj}}}{D}
  =\left(1-\left\langle\frac{h_x^2}{g}\right\rangle\right)
  =\frac{1}{2}\left(1+\left\langle\frac{1}{g}\right\rangle\right).
  \label{eq:D_proj}
\end{equation}
This relation has been derived previously by a slightly different
approach~\cite{Gustaffson} but not evaluated.
  
We evaluate the rescaling factor $D_{\text{proj}}/D$ using the classical
Helfrich Hamiltonian $\mathcal{H}_0$ as a model for the membrane. It takes the
following approximate form in the Monge representation
($\mathbf{r}\equiv(x,y)$, $\nabla\equiv(\partial/\partial x,\partial/\partial
y)$)
\begin{equation}
  \mathcal{H}_0[h(\mathbf{r},t)]=\int_A\!\mathrm{d}^2\mathbf{r}\,
  \frac{\kappa}{2}(\nabla^2 h)^2+\frac{\sigma}{2}(\nabla h)^2,
  \label{eq:Helf_Ham}
\end{equation} 
where $\kappa$ is the bending rigidity of the membrane and $\sigma$ an
effective tension~\cite{Seifert:1997}.  To calculate the expression $\langle
1/g\rangle= \int_0^{\infty}\!\mathrm{d}\alpha\,\langle\exp[-\alpha g]\rangle$ we
introduce the functional integral
\begin{equation}
  \tilde{Z}(\alpha)=\int\!\mathcal{D}[h]\,
  \exp\left[-\beta\int_{L^2}\!\mathrm{d}^2\mathbf{r}\,\frac{\kappa}{2}\left(\nabla^2
      h\right)^2
    +\left(\frac{\sigma}{2}+\frac{\alpha}{\beta L^2}\right)(\nabla h)^2\right],
\end{equation}
where $L$ is the length of the system in the $x$- and $y$-direction and
$\beta\equiv 1/(k_BT)$ the inverse temperature.  The partition function $Z$ of
the membrane is given by $\tilde{Z}(\alpha=0)$ and therefore $\langle
\exp[-\alpha(\nabla h)^2]\rangle=\tilde{Z}(\alpha)/\tilde{Z}(0)$.  Using a
Fourier expansion of spatially varying variables the calculation of
$\tilde{Z}(\alpha)$ is straightforward and yields
\begin{equation}
  \langle
  \exp[-\alpha(\nabla h)^2]\rangle
  =\exp\left[-\frac{1}{2}\sum_{\genfrac{}{}{0pt}{}{k_x,k_y}
      {|\mathbf{k}|<q_m}}\ln\left(1+\frac{2\alpha}{\beta L^2}
      \frac{1}{\kappa k^2+\sigma}\right)\right].
\end{equation}
The cutoff wave number $q_m\sim 1/a$ is given by the smallest length scale $a$
present in the membrane. Because $Lq_m\gg 1$ we replace the summation on the
rhs by the integration over $\mathbf{k}$. The ratio of projected and
intramembrane diffusion constant may then be written as
\begin{multline}
  \frac{D_{\text{proj}}}{D}
  =\frac{1}{2}+\frac{1}{2}\int_0^\infty\!\mathrm{d}\alpha\,
  \exp\left[-\alpha-\frac{1}{8\pi}\left\{\frac{\beta\sigma/q_m^2\,(Lq_m)^2}{\beta\kappa}
      \ln\left(\frac{\beta\sigma/q_m^2}{\beta\kappa+\beta\sigma/q_m^2}\right)+
      (Lq_m)^2\times\right.\right.\\
  \left.\left.\ln\left(1+\frac{2\alpha/(Lq_m)^2}{\beta\kappa
          +\beta\sigma/q_m^2}\right)
      +\frac{(Lq_m)^2}{\beta\kappa}\left(\frac{2\alpha}{(Lq_m)^2}+\beta\sigma/q_m^2\right)
      \ln\left(1+\frac{\beta\kappa}{2\alpha/(Lq_m)^2+\beta\sigma/q_m^2}\right)\right\}\right].
  \label{eq:1/g_mitt}
\end{multline}
This integral is a function of three dimensionless quantities $L q_m$,
$\beta\kappa$, and $\beta\sigma/q_m^2$.  An analytic solution is not found,
but the fast decay of the integrand facilitates the numerical evaluation.

For a numerical calculation of the scaling factor of the diffusion constant,
we first need to analyse the size of the parameters $\beta$, $\kappa$,
$\sigma$, $L$, $q_m$ which go into eq.~\eqref{eq:1/g_mitt}. Typical values of
$\beta\kappa$ for lipid bilayer membranes lie in the range between 5 and 50.
The smallest length scale $a$ of the system, that gives $q_m\sim1/a$, is
roughly the size of a lipid which is on the order of nanometres. The area
$L^2$ of membranes studied in experiments can vary strongly from a few
$\ensuremath{\unskip\,\mu \text{m}^2}$ to approximately $100\,000 \ensuremath{\unskip\,\mu \text{m}^2}$. For $(Lq_m)^2$
we, therefore, regard the range from $10^6$ to $10^{11}$. Typical values for
the effective tension of a fluctuating membrane at room temperature are
$10^{-6}\ensuremath{\unskip\,\text{mJ/m}^2}\lesssim\sigma\lesssim 10^{-3}\ensuremath{\unskip\,\text{mJ/m}^2}$, while rupture
occurs for tensions on the order of $\sigma\sim
1\ensuremath{\unskip\,\text{mJ/m}^2}$~\cite{Seifert:1997}. The whole range we regard is
$10^{-7}\leqslant \beta\sigma/q_m^2 \leqslant 10^{-1}$.

\begin{figure}
\begin{minipage}{0.45\textwidth}
\includegraphics[width=\textwidth]{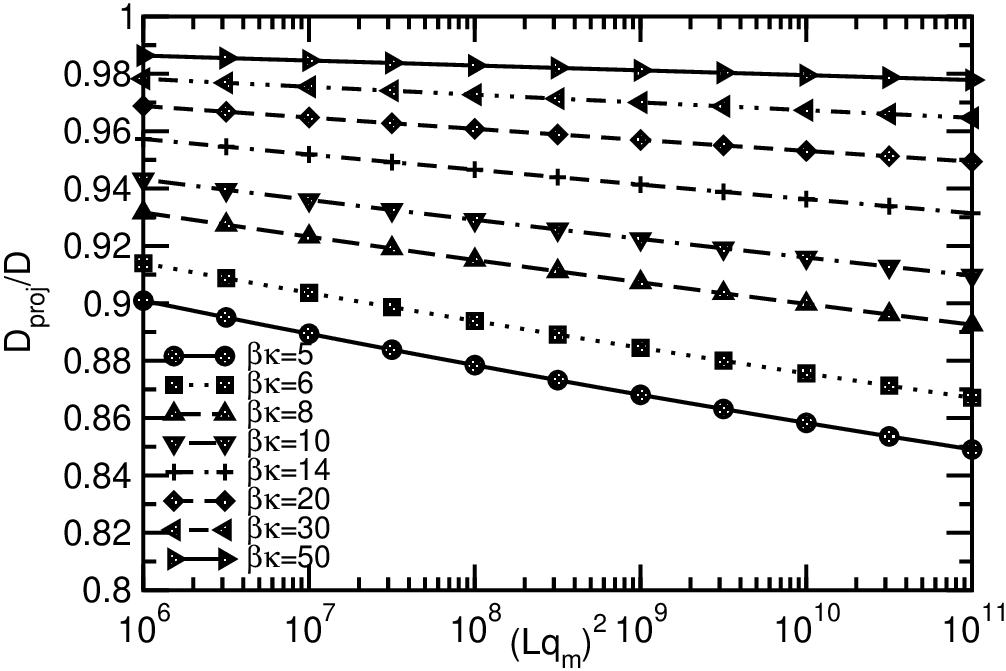}
\caption{\label{fig:Dp/D_bk_const} Rescaling factor $D_{\text{proj}}/D$ as a
    function of $(Lq_m)^2$ for the different given values of $\beta\kappa$ and
    $\sigma=0$. The softer the membrane, the stronger is the rescaling of the
    diffusion constant. The larger the regarded membrane, the stronger is the
    shift in $D$.}
\end{minipage}
\hfill
\begin{minipage}{0.45\textwidth}[t]
\includegraphics[width=\textwidth]{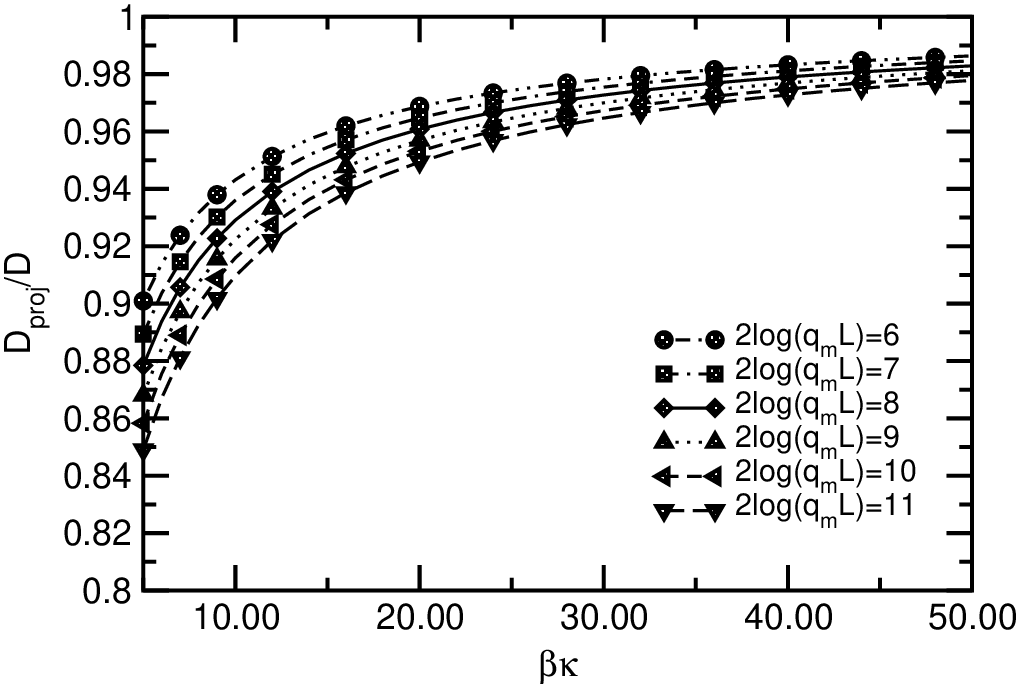}\vspace*{-0.1cm}
  \caption{\label{D_D0_kl10_5-11_bk}$D_{\text{proj}}/D$ as a
    function of rigidity $\beta\kappa$ for the given values of $(Lq_m)^2$.}
\vspace*{0.95cm}
\end{minipage}
\end{figure}
First, we analyse the case of vanishing effective tension $\sigma=0$: In
fig.~\ref{fig:Dp/D_bk_const} we show $D_{\text{proj}}/D$ as a function of
$(Lq_m)^2$ for the given values of $\beta\kappa$.  Overall we see that in the
parameter ranges, which correspond to experimental conditions, the projected
diffusion constant is reduced by up to $\sim$15\%. As was to be expected we
also find that a more rigid membrane leads to weaker rescaling of the
diffusion constant and that an increase in membrane size, which leads to
stronger fluctuations, causes the scaling factor to decrease.  In
fig.~\ref{D_D0_kl10_5-11_bk} we display $D_{\text{proj}}/D$ as a function of
rigidity $\beta\kappa$ for the given $(Lq_m)^2$. We find that an increase in
$\beta\kappa$ from 5 to 50 increases $D_{\text{proj}}/D$ monotonically by
approximately 0.1. For larger membranes, when the role of the fluctuations is
more pronounced, $D_{\text{proj}}/D$ is smaller and the rise with
$\beta\kappa$ stronger.

We now consider the influence of an effective tension. The larger $\sigma$ the
more expensive it is for the membrane to fluctuate, i.e.~to build up gradients
$|\nabla h|$. As a consequence, an increase in $\sigma$ will lead to a weaker
reduction of $D_{\text{proj}}/D$.  This can be seen in
fig.~\ref{fig:D_D0_ql2_10_7}, where we display $D_{\text{proj}}/D$ as a
function of $\beta\kappa$ and $\beta\sigma/q_m^2$ for $(Lq_m)^2=10^7$.
For the smallest regarded effective tension and bending rigidity the reduction
of the diffusion constant is the strongest with $\sim$10\%. An increase in
$\beta\kappa$ and $\beta\sigma/q_m^2$ leads to a decreased difference between
the projected and actual diffusion constant. For larger effective tensions and
smaller rigidity the lines of constant $D_{\text{proj}}/D$ appear almost
linear. This behaviour can be extracted analytically from
eq.~\eqref{eq:1/g_mitt} in the limit of large systems
\begin{equation}
  D_{\text{proj}}/D\approx
  \left\{1+\left[1+\ln\left(1+\beta\kappa/(\beta\sigma/q_m^2)\right)
      /(4\pi\,\beta\kappa)\right]^{-1}\right\}/2,
  \label{eq:Dp_D_large}
\end{equation}
which may be a more convenient expression for future reference than the full
expression~\eqref{eq:1/g_mitt}.
\section*{Curvature-coupled diffusion}
\begin{figure}
  \psfrag{bk}{\footnotesize $\beta\kappa$}
  \psfrag{bsqm}{\footnotesize $\beta\sigma/q_m^2$}
\begin{minipage}{0.45\textwidth}
\includegraphics[width=\textwidth,clip,angle=0]{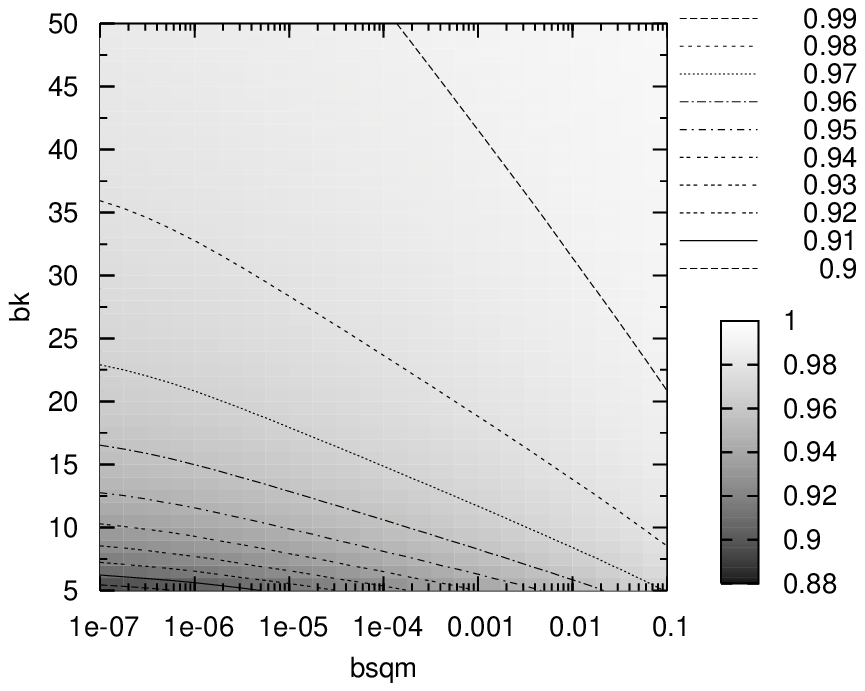}
\caption{\label{fig:D_D0_ql2_10_7} Factor $D_{\text{proj}}/D$ as a function of
    $\beta\kappa$  and $\beta\sigma/q_m^2$ for $(q_mL)^2=10^7$. The reduction in
    diffusion constant is the weakest for stiff membranes at high effective
    tension.} 
\end{minipage}
\hfill
\begin{minipage}{0.45\textwidth}
\includegraphics[width=\textwidth,clip,angle=0]{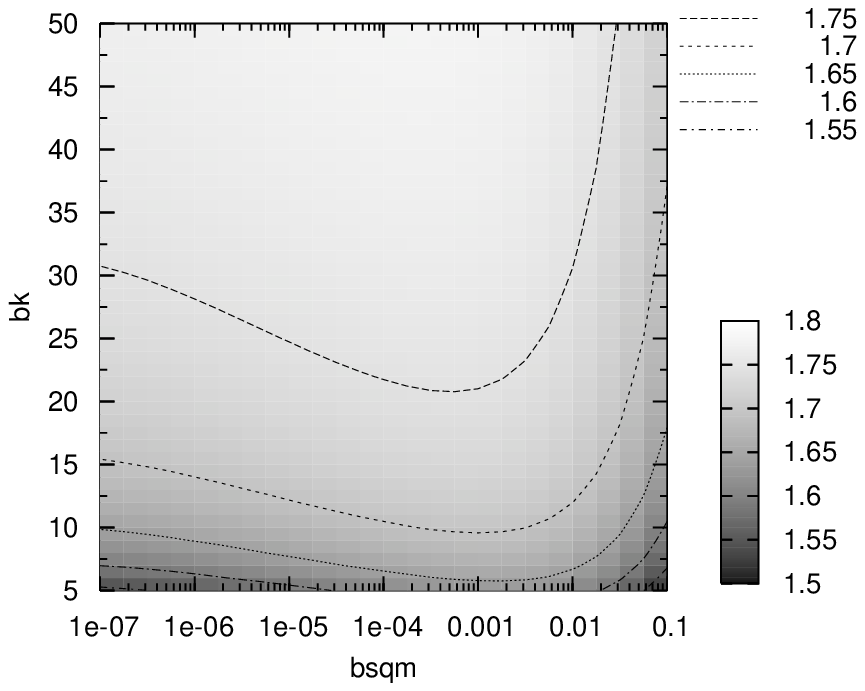}
  \caption{\label{fig:D_i_D0_qml10_7}Ratio $D_{\text{cc}}/D$ of the diffusion constant of
    a curvature-coupled protein and intramembrane diffusion as a function of
    rigidity $\beta\kappa$ and effective tension $\beta\sigma/q_m^2$ for
    $(Lq_m)^2=10^7$.}
\end{minipage}
\end{figure}
In this section we regard a protein that interacts with the fluctuating
membrane. The exact mechanisms, how a protein couples to a bilayer membrane
are not understood, however there are a large number of possibilities that
have been explored in theoretical calculations and
simulations~\cite{Weikl_2,Kim,Netz,Goulian,Dan,Lipowsky}.
In this study we model a protein of radius $a_p$ at position
$\mathbf{R}\equiv(X,Y)$ in the projected plane to couple to a local curvature
$C_p$ of the membrane via the elastic coefficient $m$. This situation could be
caused either by the shape of the protein alone, or by a protein with a large
extramembrane domain, that induces a spontaneous curvature in
the membrane~\cite{Lipowsky}. 
The energy $\mathcal{H}_1$ of this interaction,
\begin{equation}
  \mathcal{H}_1[h(\mathbf{R})]=
  \frac{m}{2}\left(\nabla^2_{\mathbf{R}}h(\mathbf{R})-C_p\right)^2-
  \frac{\kappa}{2}\pi a_p^2\left(\nabla_{\mathbf{R}}^2h(\mathbf{R})\right)^2,
\label{eq:H1}
\end{equation}
needs to be added to the membrane energy $\mathcal{H}_0$ of
eq.~\eqref{eq:Helf_Ham} to give the full energy
$\mathcal{H}=\mathcal{H}_0+\mathcal{H}_1$ of the system.  The second term of
eq.~\eqref{eq:H1} accounts for the fact that there is no membrane where the
protein is.  If we are interested in the dynamics of the system we need to
take the equations of motion both for the membrane and the particle into
account. The dynamics of the membrane is expressed by
\begin{equation}
  \partial h(\mathbf{r})/\partial t=
  -\int\!\mathrm{d}^2\mathbf{r}'\Lambda(\mathbf{r},\mathbf{r}')\,(\delta\mathcal{H}_0/\delta
  h(\mathbf{r}')+\delta\mathcal{H}_1/\delta
  h(\mathbf{r}'))+\xi(\mathbf{r})
  \label{eq:dyn_memb}
\end{equation}
with the kinetic or Onsager coefficient $\Lambda(\mathbf{r},\mathbf{r}')$ and
random fluctuations $\xi$, which obey the fluctuation dissipation theorem.  In
Fourier space, the kinetic coefficient for a free, on average planar, membrane
embedded in infinite space is given by $\Lambda(\mathbf{k})=(4\eta k)^{-1}$
with the viscosity $\eta$ of the fluid surrounding the
membrane~\cite{Seifert:1997}.  The dynamics of the inclusion follows the
Langevin equation
\begin{equation}
  \begin{split}
    \partial\mathbf{R}/\partial t&=
    -\mu_{\text{proj}}\nabla_{\mathbf{R}}\mathcal{H}+\bm{\zeta}=
    \mu_{\text{proj}}\left[m C_p+(\kappa\pi
      a_p^2-m)\nabla_{\mathbf{R}}^2h(\mathbf{R})\right]
    \nabla_{\mathbf{R}}\left(\nabla_{\mathbf{R}}^2h(\mathbf{R})\right)+\bm{\zeta},
  \end{split}
  \label{eq:dyn_part}
\end{equation}
where $\bm{\zeta}$ is a random force acting on the inclusion with
\begin{eqnarray}
  \langle\bm{\zeta}(t)\rangle&=0\quad\quad\text{and}\quad\quad\langle\zeta_i(t)\zeta_j(t')\rangle&=2
  D_{\text{proj}}\delta_{ij}\delta(t-t'),
\label{eq:rand_fluct}
\end{eqnarray}
while $\mu_{\text{proj}}$ is a mobility, that is connected to the diffusion
coefficient via the Einstein relation $k_BT\mu_{\text{proj}}=D_{\text{proj}}$.
Because we are only regarding the movement of the particle in the projected
plane we need to use the projected diffusion constant $D_{\text{proj}}$, which
we previously related to the intramembrane diffusion constant $D$.  The
solution of the coupled eqs.~\eqref{eq:dyn_memb} and \eqref{eq:dyn_part}
defines the dynamics of the system.  If we regard eq.~\eqref{eq:dyn_memb}, it
is clear that the solution will have two additive parts: the first is the
result $h_0(\mathbf{r},t)$ of the membrane without inclusion, while the second
$h_1(\mathbf{r},t)$ is a correction caused by the protein-membrane interaction
energy $\mathcal{H}_1$. To calculate the full dynamics of the particle
$h_0+h_1$ is plugged into eq.~\eqref{eq:dyn_part}.  If, however, the
interaction energy is much smaller than the unperturbed membrane energy,
$\mathcal{H}_1\ll\mathcal{H}_0$, we may use the first order approximation of
the Langevin equation~\eqref{eq:dyn_part}, which corresponds to using only the
dynamics $h_0$ of an unperturbed membrane to calculate the protein dynamics.

This approximate form of eq.~\eqref{eq:dyn_part} is now used to calculate the
diffusion constant of the curvature-coupled protein inclusion defined as
$D_{\text{cc}}\equiv\lim_{t\to\infty}\langle\Delta\mathbf{R}^2\rangle/4t$.
$\Delta\mathbf{R}^2$ is the projected squared distance in the $X$-$Y$-plane
the particle has moved during time $t$. In the following we will drop the
subscript of $h$ and assume that it is the result of the unperturbed membrane.
Using eq.~\eqref{eq:dyn_part} we write ($\mathbf{R}(t=0)=0$)
\begin{eqnarray} 
  \langle\Delta\mathbf{R}^2(t)\rangle=
  &\displaystyle\int_0^t\!\int_0^t&\!\mathrm{d}\tau\,\mathrm{d}\tau'
  \left\langle\frac{\partial\mathbf{R}(\tau)}{\partial\tau}
    \frac{\partial\mathbf{R}(\tau')}{\partial\tau'}\right\rangle\nonumber\\
  =&\displaystyle\int_0^t\!\int_0^t&\!\mathrm{d}\tau\,\mathrm{d}\tau'
  \mu_{\text{proj}}^2\left\{m^2C_p^2\left\langle[\nabla(\nabla^2h(\tau))]
      [\nabla(\nabla^2h(\tau'))]\right\rangle\right.+\mathcal{O}(h^4).
  \label{eq:dR2_mittel}
\end{eqnarray}
For the last line we used eq.~\eqref{eq:rand_fluct} and remembered that
averages of uneven powers of $h$ vanish.  Because contributions on the order
of $\mathcal{O}(h^4)$ are negligible, they are omitted in the following. To
calculate the diffusion coefficient, we need the thermal average $\left\langle
  M\right\rangle\equiv\left\langle[\nabla(\nabla^2h(t))]
  [\nabla(\nabla^2h(0))]\right\rangle$.  We introduce the function
$\mathcal{Z}(\alpha)$
\begin{equation}
  \mathcal{Z}(\alpha)=\int\mathcal{D}[h]\exp\left[-\beta\frac{1}{(2\pi)^2}\int\!
    \mathrm{d}^2\mathbf{k}\left\{\frac{E(k)}{2}-\frac{\alpha}{\beta
        L^2}k^6\exp\left[-\Lambda(k)E(k)t\right]\right\}h(k)h^*(k)\right],
  \label{eq:mathcal_Z}
\end{equation}
with $E(k)\equiv\kappa k^4+\sigma k^2$.  The desired quantity is then
determined by $\langle
M\rangle=\frac{1}{\mathcal{Z}(0)}\frac{\partial\mathcal{Z}(0)}{\partial\alpha}$.
Inserting the evaluation of eq.~\eqref{eq:mathcal_Z} into
eq.~\eqref{eq:dR2_mittel} and performing the two time integrals yields
\begin{equation}
  \left\langle\Delta\mathbf{R}^2(t)\right\rangle=4D_{\text{proj}}\,
  t+\mu_{\text{proj}}^2m^2C_p^2\frac{1}{2\pi}\int_{0}^{q_{m}}\!\mathrm{d} k
  \frac{2k^7}{\beta E^2(k)\Lambda(k)}\left\{t+\frac{\exp[-\Lambda(k)E(k)t]-1}{\Lambda(k)E(k)}\right\},
\end{equation}
from which we derive the diffusion constant $D_{\text{cc}}$ for
curvature-coupling as
\begin{equation}
  \begin{split}
    D_{\text{cc}} &=D_{\text{proj}}\left\{1+\mu_{\text{proj}}\,
      m^2C_p^2\,\frac{1}{\pi}\,\eta\,\frac{q_{m}}{\kappa^2}\left[1+\frac{1}{2\left(1+\frac{\kappa}{\sigma}
            q_{m}^2\right)}-
        \frac{3}{2}\frac{1}{\sqrt{\frac{\kappa}{\sigma}}q_{m}}
        \arctan\left(\sqrt{\frac{\kappa}{\sigma}}q_{m}\right)\right]\right\}.
  \end{split}
  \label{eq:D_i}
\end{equation}
Contrary to the last section, where we could give quantitative predictions for
the relation between projected and free intramembrane diffusion, this
expression is subject to more model parameters and is less universal.
However, it is still instructive to make a semi-quantitative estimate to gain
insight into the influence of curvature-coupled protein diffusion.

The factor $[\ldots]$ on the right of eq.~\eqref{eq:D_i}, that contains only
$\sqrt{\kappa/\sigma}q_m^2$, is always very close to one. Only towards the low
end of the regarded range $10\leqslant\sqrt{\kappa/\sigma}q_m^2\leqslant 10^4$
is this factor slightly reduced. Therefore, the prefactor of $[\ldots]$, which
contains the elastic constant $m$, the spontaneous curvature $C_p$, the
rigidity $\kappa$, the cutoff wave number $q_m$, the viscosity $\eta$, and the
mobility $\mu_{\text{proj}}$, determines the magnitude of the diffusion.
Because this prefactor is obviously positive, the diffusion of a protein is
generally enhanced for the considered coupling to local curvature.

The characteristic scale of the elastic constant $m$ is the bending rigidity
$\kappa$ of the membrane times the area $\pi a_p^2$ occupied by the protein.
Thus we can write $m=c_1\,\kappa\,\pi a_p^2$ with a constant $c_1$ on order of
unity. We choose $c_1=2$. An axisymmetric inclusion, whose area on one side of
a membrane with width $d$ is twice that of the other side, produces a
spontaneous curvature of $C_p=4/3d$~\cite{Seifert:1997}. A good estimate for
the cutoff wave number is $q_{m}=\frac{\pi}{2a}$. We further assume the radius
of the protein to be $a_p=8a$, and the width of the membrane $d=5a$. These
estimates ensure that the energy of the protein may still be regarded as a
perturbation.  Unlike the initial problem, where we could give the ratio
$D_{\text{proj}}/D$, we now also need an estimate for the mobility
$\mu_{\text{proj}}=\beta D_{\text{proj}}$.  Saffman and Delbr\"uck
\cite{Saffman:1975} derived this expression for the mobility within a flat
two-dimensional liquid layer bound by a surrounding three-dimensional fluid
with viscosity $\eta$
\begin{equation}
  \mu'=\left[\ln\left(\nu d/(\eta
  a_p)\right)-\gamma\right]/(4\pi\nu d).
  \label{eq:Saffmann}
\end{equation}
In this equation, $\nu$ is the viscosity of the liquid layer, i.e.~the
membrane, and $\gamma\simeq 0.577$ is the Euler number. This mobility is valid
for $\nu d/\eta a_p >>1$, which is the case for a typical lipid membrane. We
use eq.~\eqref{eq:Saffmann} with the previously calculated rescaling,
$\mu_{\text{proj}}=\mu'D_{\text{proj}}/D$, for the estimate of the diffusion
constant $D_{\text{cc}}$.  For a lipid membrane, the viscosity is on the order
of $\nu\simeq 1\text{erg sec/cm}^3$ and the viscosity of water is $\eta\simeq
10^{-2}\text{erg sec/cm}^3$. This gives $\eta/\nu\simeq 0.01$. If we neglect
the effective tension and use our estimates in eq.~\eqref{eq:D_i}, we find:
\begin{equation}
  D_{\text{cc}}(\sigma=0)/D\simeq
  D_{\text{proj}}/D\left\{1+0.81\,D_{\text{proj}}/D\right\}
\end{equation}
For a typical ratio of $D_{\text{proj}}/D=0.95$ we see that the diffusion of
the protein is increased by~$\sim$68\%. This estimate reveals that a coupling
to local curvature can lead to a significantly enhanced diffusion coefficient.
                           
The ratio $D_{\text{cc}}/D$ for $\sigma>0$ using $D_{\text{proj}}/D$ from
eq.~\eqref{eq:1/g_mitt} is shown in fig.~\ref{fig:D_i_D0_qml10_7} as a
function of $\beta\sigma/q_m^2$ and $\beta\kappa$ for $(Lq_m)^2=10^7$.
Surprisingly the weakest influence on the diffusion constant is found for
small rigidity $\beta\kappa$ and small effective tension $\beta\sigma/q_m^2$,
i.e.~for membranes with strong fluctuations. This follows from a partial
compensation of the two effects discussed in this paper: while the
curvature-coupled interaction between membrane and protein enhances diffusion,
the fluctuations reduce the projected diffusion in the $x$-$y$-plane,
cf.~fig.~\ref{fig:D_D0_ql2_10_7}. Starting from small rigidity and tension we
see that the diffusion coefficient increases when the rigidity and effective
tension are raised. Increasing $\beta\sigma/q_m^2$ and $\beta\kappa$ leads to
weaker fluctuations in the membrane, which finally becomes unfavourable for
the protein-membrane interaction. This becomes especially visible, in the case
of constant $\beta\kappa$ and increasing $\beta\sigma/q_m^2$: initially, an
increase in the diffusion constant is observed followed by a subsequent
decrease.
\section*{Concluding perspective}
In summary, we have found that thermal fluctuations of biomembranes have
considerable influence on the diffusion of proteins.  When the protein is free
to diffuse within the membrane, the projected diffusion constant, which
corresponds to the quantity typically measured, is up to 15\% smaller than
that of the true intramembrane diffusion. The lower the bending rigidity or
the effective tension and the larger the membrane, the stronger is this
effect. Experimentally, it could be studied by observing the change in the
projected diffusion constant upon osmotically induced swelling of an initially
flaccid vesicle, which increases the effective tension~\cite{Seifert:1997}.
To gain insight into the influence of protein-membrane interactions, we
studied the diffusion of an inclusion with a spontaneous curvature and found
that this enhances diffusion significantly.  When the projected diffusion
constant is measured this leads to an interesting interplay: on the one hand
fluctuations can lead to enhanced intramembrane diffusion, while on the other
hand stronger fluctuations lead to a smaller projected diffusion constant. It
is therefore possible that the measurement of the diffusion coefficient at
constant rigidity and increasing effective tension reveals a maximum for a
certain effective tension.  Future experimental studies should analyse the
effect of changing membrane fluctuations on the lateral diffusion of proteins,
since they could shed light on the local coupling mechanisms between proteins
and lipids.

\end{document}